\documentclass{vgtc}                          

\graphicspath{{figures/}{pictures/}{images/}{./}} 

\usepackage{times}                     

\usepackage{tabu}                      
\usepackage{booktabs}                  
\usepackage{lipsum}                    
\usepackage{mwe}                       
\usepackage{amsmath}
\usepackage{colortbl}
\usepackage{amssymb}
\usepackage{soul}
\usepackage[subrefformat=parens,labelformat=parens]{subfig}
\usepackage{pifont}
\newcommand{\cmark}{\ding{51}}%
\newcommand{\xmark}{\ding{55}}%
\usepackage{soul}
\usepackage{bm}

\usepackage{mathptmx}                  

\usepackage{multirow,multicol}
\usepackage{array}
\newcolumntype{x}[1]{>{\centering\arraybackslash\hspace{0pt}}p{#1}}
\onlineid{0}

\vgtccategory{Research}

\vgtcinsertpkg
\usepackage{tikz}
\usepackage{pgfplots}
\usetikzlibrary{pgfplots.colormaps}
\pgfplotsset{compat=1.16}
\usepackage{pgfplotstable}\newcommand*{\ReadOutElement}[4]{%
    \pgfplotstablegetelem{#2}{[index]#3}\of{#1}%
    \let#4\pgfplotsretval
}

\definecolor{limegreen}{RGB}{50,205,50}

\usepackage{glossaries}
\newacronym{cnn}{CNN}{Convolutional Neural Network}
\newacronym{dof}{DoF}{Degrees of Freedom}
\newacronym{gru}{GRU}{Gated Recurrent Units}
\newacronym{ml}{ML}{machine learning}
\newacronym{mlp}{MLP}{Multi-Layer Perceptron}
\newacronym{hci}{HCI}{Human-Computer Interaction}
\newacronym{hmd}{HMD}{Head Mounted Display}
\newacronym{knn}{KNN}{K-Nearest Neighbors}
\newacronym{lstm}{LSTM}{Long-Short Term Memory}
\newacronym{vr}{VR}{Virtual Reality}
\newacronym{idc}{IDC}{International Data Corporation}
\newacronym{lgbm}{LGBM}{Lightweight Gradient Boosting Model}
\newacronym{rf}{RF}{Random Forest}
\newacronym{qda}{QDA}{Quadratic Discriminant Analysis}
\newacronym{lr}{LR}{Logistic Regression}
\newacronym{qos}{QoS}{Quality of Service}

\definecolor{color1}{HTML}{440154}
\definecolor{color2}{HTML}{31688e}
\definecolor{color3}{HTML}{35b779}
\definecolor{color4}{HTML}{ffa600}
\newcommand{\edit}[1]{\textcolor{black}{#1}}

\title{Movement- and Traffic-based User Identification in Commercial Virtual Reality Applications: Threats and Opportunities}

\author{Sara Baldoni, Salim Benhamadi,
Federico Chiariotti, Michele Zorzi, and
Federica Battisti} %
\affiliation{Department of Information Engineering, University of Padova, Padua, Italy}

\teaser{
  \centering
  \includegraphics[width=\linewidth]{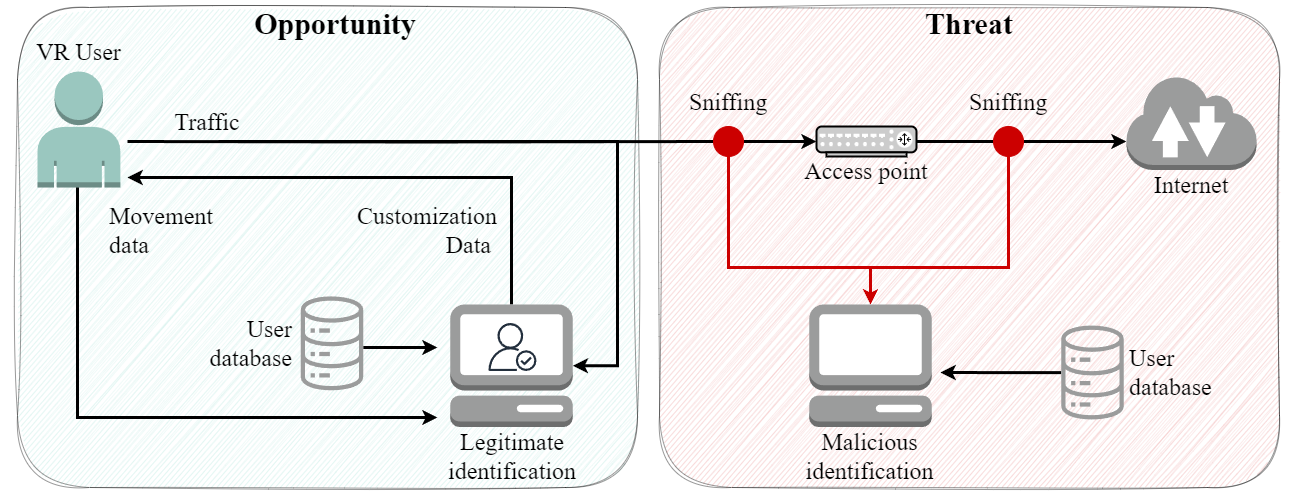}
  \caption{Overview of threats and opportunities in network-based VR applications.}
  \label{fig:teaser}
}

\glsdisablehyper

\abstract{
    With the unprecedented diffusion of virtual reality, the number of application scenarios is continuously growing. As commercial and gaming applications become pervasive, the need for the secure and convenient identification of users, often overlooked by the research in immersive media, is becoming more and more pressing. Networked scenarios such as Cloud gaming or cooperative virtual training and teleoperation require both a user-friendly and streamlined experience and user privacy and security. In this work, we investigate the possibility of identifying users from their movement patterns and data traffic traces while playing four commercial games, using a publicly available dataset. If, on the one hand, this paves the way for easy identification and automatic customization of the virtual reality content, it also represents a serious threat to users' privacy due to network analysis-based fingerprinting. Based on this, we analyze the threats and opportunities for virtual reality users' security and privacy.
} 


\begin{document}


\firstsection{Introduction}

\maketitle
\glsresetall

In the last years, there has been a significant rise in the adoption of \gls{vr} technology. According to a recent analysis performed by the \gls{idc} and published in \cite{idc_24}, the shipment of \gls{vr} headsets is expected to reach $24.7$~million units by 2029, marking a $29.2\%$ growth over the past five years.
A significant booster is the expansion of use cases for this technology: beyond gaming, \gls{vr} is now widely adopted for training, teaching, and healthcare. While much of the scientific community's effort is focused on developing efficient methods and algorithms for content creation, delivery, and storage, the area of user identification for device access remains underexplored. 

Identifying the user of a \gls{vr} device can be useful in many scenarios, especially in a multi-user environment where the same \gls{hmd} can be shared among several users. In this context, user identification can be exploited to customize the \gls{vr} experience by loading personal settings and preferences.
The most widely adopted strategy for \gls{vr} user authentication is through passwords or PIN codes that are typically inserted in the \gls{vr} environment through the controllers. This action inherently poses security challenges, as the password or PIN pattern may be recorded, recognized, or stolen by a malicious user \cite{George_2017_NDSS}. \edit{Therefore, movement patterns, which are less predictable and repeatable by an intruder, have been proposed as alternative identification source.}

It is also worth mentioning that, as \gls{vr} experiences become more realistic, the complexity of rendering the virtual scenes is growing beyond the capabilities of compact \glspl{hmd}, whose computational power, cooling capabilities, and battery life are inherently limited by their weight and size. For this reason, \gls{vr} rendering is moving to the Cloud~\cite{mehrabi2021multi}. However, the networking challenges posed by the tight \gls{qos} constraints~\cite{zhao2021virtual} often require the use of predictive compression and resource allocation techniques~\cite{chiariotti2023temporal}. Since these techniques can be based on the specific actions and movements of the user, we can consider the \edit{amount of} data traffic exchanged between the rendering server and the headset as a behavioral fingerprint. \edit{Therefore, we decided to complement the movement data with information about the amount of \gls{vr} network traffic, which, to the best of our knowledge, has never been employed in this specific application.}

\begin{table*}[htb]
\caption{Summary of existing user identification approaches in the \gls{vr} literature.}
\centering
\renewcommand*{\arraystretch}{1.05}
\begin{tabular}{c|ccc|c|c|c|c}
\toprule
\multicolumn{1}{c|}{\multirow{2}{*}{Approach}} & \multicolumn{3}{c|}{Body features}   &  \multicolumn{1}{c|}{\multirow{2}{*}{Traffic}}  & \multicolumn{1}{c|}{\multirow{2}{*}{Ad hoc task}} & \multicolumn{1}{c|}{\multirow{2}{*}{Repeated sessions}} & \multirow{2}{*}{Number of users}  \\
\multicolumn{1}{c|}{}                          & Eye & Hands & \multicolumn{1}{c|}{Head} & \multicolumn{1}{c|}{}                             & \multicolumn{1}{c|}{}                                   &                              \\ \midrule
                    Kupin \emph{et al.} \cite{Kupin_MM_2019}                           &    \textcolor{red}{\xmark}             &   \textcolor{limegreen}{\cmark}   &      \textcolor{red}{\xmark}         & \textcolor{red}{\xmark}             &    \textcolor{limegreen}{\cmark}                                               &            \textcolor{limegreen}{\cmark}                                             &        14                      \\ \hline
                    Pfeuffer \emph{et al.} \cite{Pfeuffer_CHI_2019}                           &  \textcolor{limegreen}{\cmark}               &   \textcolor{limegreen}{\cmark}    &   \textcolor{limegreen}{\cmark}    & \textcolor{red}{\xmark}                     & \textcolor{limegreen}{\cmark}                                                   &     \textcolor{limegreen}{\cmark}                                                     &            22       
                    \\ \hline
                    Olade \emph{et al.} \cite{Olade_Sensors_2020}                           &    \textcolor{limegreen}{\cmark}              &   \textcolor{limegreen}{\cmark}    &       \textcolor{limegreen}{\cmark}        &\textcolor{red}{\xmark}              &      \textcolor{limegreen}{\cmark}                                              &     \textcolor{limegreen}{\cmark}                                                    &     15
                    \\ \hline
                    Liebers \emph{et al.} \cite{Liebers_CHI_2021}                           &  \textcolor{red}{\xmark}              &   \textcolor{limegreen}{\cmark}    &         \textcolor{limegreen}{\cmark}        &    \textcolor{red}{\xmark}          &                                            \textcolor{limegreen}{\cmark}        &   \textcolor{limegreen}{\cmark}                                                      &   16  
                    \\ \hline
                    Liebers \emph{et al.} \cite{Liebers_VRST_2023}                           &  \textcolor{red}{\xmark}                  &   \textcolor{limegreen}{\cmark}     &             \textcolor{limegreen}{\cmark}      &    \textcolor{red}{\xmark}         &    \textcolor{red}{\xmark}                                                 &       \textcolor{limegreen}{\cmark}                                                     & 15 \\   \hline
                    Asish \emph{et al.} \cite{Asish_VW_2022}                           &        \textcolor{limegreen}{\cmark}        &   \textcolor{red}{\xmark}    &          \textcolor{red}{\xmark}      & \textcolor{red}{\xmark}            &    \textcolor{red}{\xmark}                                               &      \textcolor{limegreen}{\cmark}                                                   & 34 \\ \hline  
                    Wilson \emph{et al.} \cite{Wilson_TVCG_2024}                           &        \textcolor{limegreen}{\cmark}        &   \textcolor{red}{\xmark}    &          \textcolor{red}{\xmark}      & \textcolor{red}{\xmark}            &    \textcolor{limegreen}{\cmark}                                               &      \textcolor{limegreen}{\cmark}                                                   & 26 \\ \hline  
                    Rack \emph{et al.} \cite{Rack_FVR_2023}                           &        \textcolor{red}{\xmark}        &   \textcolor{limegreen}{\cmark}    &          \textcolor{limegreen}{\cmark}      & \textcolor{red}{\xmark}            &    \textcolor{red}{\xmark}                                               &      \textcolor{limegreen}{\cmark}                                                   & 71 \\ \hline
                    Rack \emph{et al.} \cite{Rack_VRST_2024}                           &        \textcolor{red}{\xmark}        &   \textcolor{limegreen}{\cmark}    &          \textcolor{limegreen}{\cmark}      & \textcolor{red}{\xmark}            &    \textcolor{limegreen}{\cmark}                                               &      \textcolor{limegreen}{\cmark}                                                   & 48 \\ \hline
                    Nair \emph{et al.} \cite{Nair_USENIX_2023}                           &        \textcolor{red}{\xmark}        &   \textcolor{limegreen}{\cmark}    &          \textcolor{limegreen}{\cmark}      & \textcolor{red}{\xmark}            &    \textcolor{red}{\xmark}                                               &      \textcolor{limegreen}{\cmark}                                                   & 55541\\
                    \bottomrule

                    \textbf{Ours}                           &  \textcolor{red}{\xmark}              &    \textcolor{limegreen}{\cmark}    &          \textcolor{limegreen}{\cmark}                  & \textcolor{limegreen}{\cmark}   &    \textcolor{red}{\xmark}                                              &         \textcolor{red}{\xmark}                                                & 60 (30 per game) \\  
                    \bottomrule
\end{tabular}\label{tab:soa}
\end{table*}

While the possibility of adding traffic features for user identification represents an opportunity to enrich the data provided to the classification engine, as shown in Figure \ref{fig:teaser}, it also opens up new and unforeseen threats. In fact, sniffing techniques on either the wireless or the wired medium may be exploited to gather traffic data for user identification. \edit{This threat has been investigated for non-\gls{vr} applications in~\cite{Bakhshandeh_ISCISC_2018}, which performs a traffic analysis at the flow level including various information such as source and destination IPs and ports, start and end time, flow duration, and amount of traffic. In~\cite{Kim_SPW_2021}, the authors propose exploiting the temporal patterns of dynamic web traffic to infer keystroke-related information. To the best of our knowledge, the extension of this threat to \gls{vr} has been only partially addressed in~\cite{Nair_UIST_2023}, mainly targeting user geolocalization. 
On the other hand, our work jointly considers traffic and movement information for performing user identification. Moreover, with respect to~\cite{Bakhshandeh_ISCISC_2018, Kim_SPW_2021}, we employ only the amount of traffic as network-related information. To the best of our knowledge, this kind of threat scenario has not been investigated in the scientific literature so far.}

In this work, we contribute to addressing these challenges by proposing a framework for user identification in \gls{vr} in which:
\begin{itemize}
    \item We perform a detailed analysis of movement and traffic data collected from $30$ users playing commercial games in \gls{vr} that are available in the Questset public dataset~\cite{baldoni2024questset};
    \item We propose a compact representation of features from movement and traffic data that can effectively be used for characterizing users;
    \item We select existing supervised learning approaches that work efficiently with a small set of training data and use them for user identification;
    \item We identify the three most significant features that best capture user behavior while playing commercial games.
\end{itemize}
Finally, we discuss the opportunities and threats of the proposed framework by evaluating the variation of the identification performance with different design options (e.g., number of users, type of input data, duration of the identification procedure).

\section{Related Work}

Over the past few years, biometrics have emerged as an effective and efficient identification and authentication method. With respect to standard techniques such as passwords and PIN codes, the usage of biometric features makes it possible for users to be recognized based on their intrinsic characteristics, without the need to remember a secret passcode~\cite{Dargan_ESA_2020}. 
Biometric identification techniques can be divided into two main branches: physiological and behavioral. While the former is based on physical features such as fingerprints, retinal patterns, and facial features, the latter aims at recognizing users based on how they perform a task, e.g., on their signature and gait~\cite{Tse_ISCAIE_2020}.
Unlike physiological methods, behavioral biometrics does not require specific hardware and enables continuous identification. Moreover, it represents a more powerful security measure with respect to physiological biometrics, since spoofing and repetition attacks require more advanced capabilities~\cite{Stragapede_PRL_2022}.

With the wider diffusion of computers and mobile phones, the concept of behavior has evolved from intrinsically human activities to include the interaction between the user and a specific technology. Keystrokes~\cite{Roy_Access_2022}, mouse patterns~\cite{Zhang_BIT_2024} and touch screen swipes~\cite{Delgado_ESA_2024} have all been employed in this direction.

The shift from traditional to immersive technologies led to a further evolution of behavioral biometrics. Differently from other \gls{hci} methods, \gls{vr} allows the users to perform actions with their whole body. Thanks to the six \gls{dof}, users can move within the virtual environment, explore it in every direction, and interact with it. Therefore, movement and eye patters play a key role in user identification. Moreover, behavioral biometrics is particularly suitable for \gls{vr}, as it makes it possible to identify users without requiring the interruption of the immersive experience.

In this direction, several works explored the possibility of performing user identification in \gls{vr} based on behavioral biometrics. These studies can be differentiated based on the type of behavioral features extracted, the definition of ad hoc identification tasks, the number of participants involved, and the test across multiple sessions. A summary of the existing approaches is provided in Table~\ref{tab:soa}.

In~\cite{Kupin_MM_2019}, a task-driven user identification in \gls{vr} has been proposed. More specifically, the study considers an ad hoc application that requires the user to throw a ball at a target. Identification is then performed by matching the user hand trajectory with a database of $14$ trajectories recorded in a different session. The presented method achieves a recognition rate of $92.86\%$.
Since \glspl{hmd} usually make it possible to record the head position, hand movements can be complemented by the head coordinates. Based on this, another task-based identification approach has been proposed in~\cite{Liebers_CHI_2021}, where $16$ users have been involved in bowling and archery games. Data were collected in two sessions, one for training and the other for testing. The authors compared the performance of a \gls{lstm} network and a \gls{mlp} of different feature sets. More specifically, they combined the information coming from the \gls{hmd}, the controllers, the considered sub-task, and the time passed from the beginning of the task. Their results show that the \gls{lstm} achieves a higher accuracy, with a maximum of $90\%$ and $68\%$ for the archery and bowling tasks, respectively. \edit{In addition, in~\cite{Rack_VRST_2024}, motion profiles have been employed to identify users while hand-writing a password in \gls{vr}. The study involved $48$ participants and two recording sessions, showing promising performance.}

Moreover, some \glspl{hmd} include an integrated eye-tracker that records eye gaze. \edit{This has been considered in~\cite{Wilson_TVCG_2024}, where the authors analyzed the identification threat rising from eye tracking. They proposed various privacy mechanisms which could be employed for reducing the identification risk.}
\edit{Gaze information has also been exploited in~\cite{Pfeuffer_CHI_2019}, which proposed an identification approach based on the combination of eye gaze and hand and head movements focusing on four tasks}. In more detail, pointing, grabbing, walking, and typing operations have been considered. The study involved $22$ users that performed the tasks multiple times in two different sessions. In addition, different target sizes and distances have been considered and two walking paths have been analyzed. A \gls{rf} classifier has been trained, achieving an accuracy of $63.55\%$ for pointing, $45.84\%$ for grabbing, $49.67\%$ for walking, and $54.27\%$ for typing. 
Similarly, another recent study~\cite{Olade_Sensors_2020} employed hand, head, and eye gaze data, considering six tasks for identification. In more detail, they designed an ad hoc application involving the grabbing, transporting, and dropping of balls and cubes. They recorded data from $15$ users in multiple sessions over two days and trained a \gls{knn} classifier achieving an accuracy of $98.60\%$.

Although task-oriented procedures may achieve high performance, they require the user to perform a specific action to complete the identification procedure. Moreover, continuous identification is hindered. In contrast, more general identification approaches have been presented \edit{in~\cite{Asish_VW_2022, Liebers_VRST_2023, Rack_FVR_2023, Nair_USENIX_2023}}. In the first paper, the authors refer to an education scenario consisting of four learning sessions acquired on the same day. They collected eye tracking data for $34$ participants and tested both machine learning (\gls{rf} and \gls{knn}) and deep learning (\gls{cnn}, \gls{lstm}) methods. When using data from all sessions together, \gls{knn} was the best performing approach, with a $99.62\%$ accuracy. The authors also evaluated the identification performance by training on three sessions and testing on the fourth. The presented results show an average accuracy of $96.09\%$ on the four sessions for \gls{knn}.
In ~\cite{Liebers_VRST_2023} a commercial game has been used for user identification. 
The authors studied the temporal stability of head and hand movements for behavioral biometrics in \gls{vr}. In more detail, they considered multiple sessions spanning across eight weeks, involving $15$ participants. An \gls{rf} classifier was trained for the identification task. 
Their results show that when training on the first session and testing on the fourth, a decrease of $38\%$ occurs in terms of the F1 score. However, they also showed that continuous training makes it possible to increase the stability of behavioral \gls{vr} data. In case of commercial applications such as games, this scenario can be considered realistic as the users will presumably play the same game more than once.
\edit{Similarly, in~\cite{Rack_FVR_2023}, the authors performed user identification for a set of $71$ users recording the \gls{hmd} and controller movements when playing a commercial game. They extracted sequences of $20$ seconds which have been processed by a \gls{cnn} and a \gls{gru}. The identification performance was evaluated using training times ranging from $1$ to $45$ minutes. Finally, in~\cite{Nair_USENIX_2023}, hand and head movement data have been complemented with game-specific contextual information for performing user identification with a large sample ($55541$ users). The study showed that, while static information (e.g., height) is enough for small user sets, motion patterns and contextual data are relevant when the number of participants increases. They achieved an accuracy of $94.33\%$ by providing a $100$-second movement sequence as input to a multi-classifier machine learning model. The authors showed that a reduction in the sequence length to 10 seconds led the identification performance to decrease to $73.2\%$.}

In this work, we consider a publicly available dataset including four commercial games to perform user identification. We adopted a feature engineering approach for various machine learning methods, including both traffic and movement data. Therefore, differently from previous approaches, we complement the movement-based biometric fingerprint with the traffic-based one.
The usage of multiple commercial applications highlights the opportunities derived from the application of behavioral biometrics to \gls{vr}. \edit{The exploitation of network traffic information for \gls{vr} privacy breaches has been partially addressed in~\cite{Nair_POPETS_2023}. The authors employed the round-trip time between
the \gls{vr} client and the server to identify the user location via multilateration, and extracted some information about the user's computer based on the bandwidth usage. Differently, we exploit the amount of \gls{vr} network traffic to perform direct user identification, so that we are the first to consider the threats from malicious actors in the network using packet sniffing as a side channel to gain information on the identity of the \gls{vr}~user.}


\section{Dataset and Machine Learning Algorithms}
This section presents the proposed identification approach by focusing on the selected dataset, the data pre-processing steps, the feature selection and engineering process, and the selected supervised learning methods.

\subsection{Dataset}

This work relies on the Questset dataset~\cite{baldoni2024questset}, which contains $120$ movement and data traffic traces from users playing commercial \gls{vr} games on a Meta Quest 2 headset. The experimental setup involved a wired connection between the headset and a computer, using the Meta Quest Link option to render the game content on the computer's graphics card and stream it to the headset.

The resulting dataset includes both traffic and movement traces, which can be used for both identification and pattern recognition: the traffic traces were provided as a list of packet sizes and timestamps, along with the direction of the packet (from the headset to the computer or vice versa). On the other hand, the movement data are significantly richer: the dataset provides traces from the native interface of the Meta Quest 2, which include the three-dimensional position of the headset and the two controllers, as well as their orientation expressed as a quaternion, with a $60$~Hz sampling rate. 

The dataset includes traces from $60$ participants, which were divided in two groups: the first group played the games \emph{Beat Saber} and \emph{Cooking Simulator}, while the second played \emph{Medal of Honor: Above and Beyond} and \emph{Forklift Simulator}. Each participant played each game for at least $10$ minutes, so while the traces have different durations, they are all at least $10$ minutes long. \edit{The participants were between 18 and 35 years old, with a balanced mix of previous gaming experience. The gender balance in the dataset skewed towards males, with 22 female participants ($36.7\%$).}

Questset additionally includes a set of participants who did not complete the full experiment due to severe cybersickness symptoms. We did not consider those users in this work, as the duration of their traces does not reach $10$ minutes in length; conscious and unconscious compensation for cybersickness might also affect the actions and movements of these users. Additionally, the order in which participants played the two games assigned to them is reported in the dataset, with half of each group playing the games in reverse order, but we did not consider these subgroups as separate in our analysis.

\subsection{Data pre-processing and feature selection}
The position and orientation of each controller and headset can be represented as a $7$-dimensional vector, i.e., the combination of $3$ spatial coordinates and an orientation quaternion. Therefore, each sample in the movement dataset has $21$ dimensions. Considering that multiple seconds of data may be needed to capture user-specific motion patterns, we selected an accumulation period of $10$~s. However, this implies that using movement data at a sampling rate of $60$~Hz for each dimension would return thousands of input samples,  making training ineffective. For this reason, we abstracted summary features to represent each dimension, reducing $600$ samples per dimension per period to just a few values.

The selected features are the average, minimum, and maximum value in the time period, along with the quartile values and the standard deviation over the period. We also included the empirical velocity and acceleration in each direction and over each orientation angle. This results in $7$ features for each dimension, computed on the data and their first and second-order differentials, with a total of $21$ values for each dimension. Moreover, we included $6$ additional dimensions, i.e., the distance and angle of the two controllers with respect to the headset and to each other, for which we did not consider velocity and acceleration. This set of features is used by the learning methods to detect user-specific patterns while reducing the number of inputs to the learning methods by an order of magnitude. Additionally, we performed an ablation study to test the robustness of the methods in case of imperfect calibration of the height of the user, as this would impact the measured height. To this end, we normalized the height values for both the head and the controllers by the mean value for each user, preserving the movement patterns while removing the effect of the initial height.

The traffic traces were processed to obtain $28$ features. In more detail, we considered the average packet size over the sampling period, the cumulative transmitted data in the same period, and the number of packets in each direction (i.e., from the computer to the headset and vice versa). For each of these we computed the average, minimum, and maximum value in the time period, along with the quartile values and the standard deviation over the period.

Before training, each feature was normalized by setting the minimum value in the dataset as $0$ and the maximum as $1$, to facilitate training. As all traces were at least $10$~minutes in length, we used the first $8$~minutes as the training set and the following $2$~ minutes as the test set. This was done to assess the robustness of the learning models' generalization, as some participants progressed through different levels (e.g., switching songs and increasing difficulty in Beat Saber), thus presenting the same users in slightly different contexts. As all users played $2$ games, we also considered the potential cross-game identification capabilities of this approach.

\subsection{Supervised learning approaches}

As the number of data samples in the training set is relatively small, while the complexity of the features is still significant even after our pre-processing, we chose to avoid deep learning and focus on basic approaches that have shown better robustness when trained with fewer data points. After testing several common approaches, we have selected the following learning methods as the best-performing ones: \gls{lr}, \gls{qda}, \gls{rf}, Extra Trees, a \gls{lgbm}, and an ensemble classifier including them all.\footnote{We used the implementations available in the \texttt{scikit-learn} Python library. The code for the full analysis is available at \url{https://github.com/signetlabdei/vr_user_id}} We detail them in the following:
\begin{table*}[tb]
  \caption{Accuracy and Macro F1 score of each learning method for the different games and input types among all $30$ users. The best individual model (within a $0.01$ margin) is marked in bold. The ensemble model results are underlined if they are better than the best individual model.}
  \label{tab:accuracy}
  \scriptsize%
	\centering%
\begin{tabular}{c|c|x{0.7cm}x{0.7cm}|x{0.7cm}x{0.7cm}|x{0.7cm}x{0.7cm}|x{0.7cm}x{0.7cm}|x{0.7cm}x{0.7cm}}
\toprule
\multirow{2}{*}{Game} & \multirow{2}{*}{\begin{tabular}[c]{@{}c@{}}Learning\\method\end{tabular}} & \multicolumn{2}{c|}{Movements} & \multicolumn{2}{c|}{Traffic} & \multicolumn{2}{c|}{\begin{tabular}[c]{@{}c@{}}Movements\\ (norm. height)\end{tabular}} & \multicolumn{2}{c|}{\begin{tabular}[c]{@{}c@{}}Movements\\ and traffic\end{tabular}}& \multicolumn{2}{c}{\begin{tabular}[c]{@{}c@{}c@{}}Movements\\ and traffic \\(norm. height)\end{tabular}} \\
& & $A$ & \edit{$F_1$} & $A$ & \edit{$F_1$}& $A$ & \edit{$F_1$}& $A$ & \edit{$F_1$}& $A$ & \edit{$F_1$}\\
\midrule
 \multirow{6}{*}{Beat Saber} & Extra Trees & \textbf{0.958} & \textbf{0.957} & 0.628& 0.596& 0.917 &0.914 & \textbf{0.986} & \textbf{0.986} & \textbf{0.972} & \textbf{0.971}  \\
 & \gls{lgbm} & 0.933 & 0.932 & \textbf{0.769} & \textbf{0.767} & \textbf{0.942}& \textbf{0.942}& 0.944& 0.943 & 0.942& 0.940   \\
 & \gls{lr} & 0.900 & 0.891 & 0.422& 0.362 &0.892 &0.884 &0.919 & 0.920&0.931 & 0.931  \\
 & \gls{qda} & 0.075 & 0.076 & 0.611 &0.592 &0.072 &0.074 & 0.111& 0.108& 0.097& 0.096   \\
 & \gls{rf} & 0.956& 0.955 &0.708 &0.686 &0.886 &0.882 & 0.981& 0.981&  0.969 & 0.969  \\ \cline{2-12}
 & \cellcolor{black!5} Ensemble & \cellcolor{black!5} 0.950  & \cellcolor{black!5} 0.950 & \cellcolor{black!5} \ul{0.828} & \cellcolor{black!5} \ul{0.823} &\cellcolor{black!5}  0.942 &\cellcolor{black!5}  0.940 &\cellcolor{black!5}  0.983& \cellcolor{black!5} 0.983& \cellcolor{black!5} \ul{0.981}& \cellcolor{black!5} \ul{0.981}  \\
                      \midrule
 \multirow{6}{*}{Medal of Honor} & Extra Trees & \textbf{0.775} &\textbf{0.759} & 0.553 & 0.527 & \textbf{0.661} & \textbf{0.632} & 0.864&0.848& 0.783&0.763   \\
 & \gls{lgbm} & 0.758& 0.748& 0.622 & 0.609 & \textbf{0.661} &\textbf{0.646} & \textbf{0.906} &\textbf{0.899} & \textbf{0.847} &\textbf{0.842}   \\
 & \gls{lr} & 0.694 & 0.685& 0.403& 0.357 & 0.639& 0.629 & 0.739 &0.733  &0.725 &0.712   \\
 & \gls{qda} & 0.083&	0.080 & \textbf{0.817} & \textbf{0.818} & 0.075&0.076 &0.050 & 0.051 &0.061 & 0.060  \\
 & \gls{rf} & \textbf{0.786}& \textbf{0.770}& 0.642& 0.626 &0.636 &0.598 &0.889 & 0.873 &0.781 & 0.757  \\ \cline{2-12}
 &\cellcolor{black!5}  Ensemble &\cellcolor{black!5} \ul{0.792}  & \cellcolor{black!5} \ul{0.779} &\cellcolor{black!5}  0.811 &\cellcolor{black!5}  0.809 & \cellcolor{black!5} \ul{0.683}&\cellcolor{black!5}\ul{0.662}&\cellcolor{black!5}0.906  & \cellcolor{black!5}0.899 &\cellcolor{black!5}0.847  &\cellcolor{black!5} 0.837   \\
 \midrule
 \multirow{6}{*}{Cooking Simulator} & Extra Trees & \textbf{0.819} & \textbf{0.796} & 0.522& 0.509 & \textbf{0.778} & \textbf{0.754} & \textbf{0.864} & \textbf{0.852} & 0.847 & 0.829  \\
 & \gls{lgbm} & 0.800 & 0.776& 0.583& 0.579 & \textbf{0.786}& \textbf{0.763}& 0.858 & 0.843 & \textbf{0.856}& \textbf{0.848}   \\
 & \gls{lr} & 0.719 & 0.700 & 0.322 &0.276 & 0.681 & 0.660 &0.775 &0.751 & 0.753 &0.729   \\
& \gls{qda} & 0.075 & 0.074& \textbf{0.839} & \textbf{0.844} & 0.103 & 0.099& 0.064 & 0.065 & 0.089 & 0.091   \\
 & \gls{rf} & 0.806 & 0.792& 0.603& 0.587 & 0.733&0.713 & \textbf{0.869}& \textbf{0.861}& \textbf{0.842}& \textbf{0.832}   \\ \cline{2-12}
 &\cellcolor{black!5} Ensemble &\cellcolor{black!5} 0.797 &\cellcolor{black!5} 0.768 &\cellcolor{black!5}0.806 & \cellcolor{black!5}0.798&\cellcolor{black!5} \ul{0.789} &\cellcolor{black!5} \ul{0.769} &\cellcolor{black!5}0.844 &\cellcolor{black!5}0.830  &\cellcolor{black!5} 0.839  & \cellcolor{black!5}0.828  \\
 \midrule
 \multirow{6}{*}{Forklift Simulator}  & Extra Trees & \textbf{0.997} & \textbf{0.997}&0.478 &0.478 & \textbf{0.992}& \textbf{0.992} & \textbf{0.997}& \textbf{0.997}& \textbf{0.994}&\textbf{0.994}   \\
 & \gls{lgbm} & 0.939 &0.939 & 0.569& 0.558& 0.917&0.915 &0.956 &0.955 & 0.967& 0.967  \\
  & \gls{lr} & 0.936 & 0.936 & 0.344&0.274 & 0.939&0.940 &0.947 &0.948 & 0.939&0.940  \\
& \gls{qda} & 0.128 & 0.135 & \textbf{0.842} & \textbf{0.832} &0.128 &0.121 &0.125 &0.127 &0.092 &0.087   \\
 & \gls{rf} & 0.975& 0.973& 0.578& 0.565& 0.972& 0.972&0.978 &0.976 &0.978&0.978   \\ \cline{2-12}
 &\cellcolor{black!5} Ensemble &\cellcolor{black!5} 0.978 &\cellcolor{black!5} 0.978 &\cellcolor{black!5} 0.775 &\cellcolor{black!5} 0.766 & \cellcolor{black!5}0.969&\cellcolor{black!5}0.969 & \cellcolor{black!5} 0.989 & \cellcolor{black!5} 0.989 &\cellcolor{black!5}0.983 &\cellcolor{black!5}0.983   \\
\bottomrule   
\end{tabular}\label{tab:results}
\end{table*}
\begin{itemize}
    \item \emph{\gls{lr}}~\cite{verhulst1847recherches,cramer2004early}: this classical method applies class-by-class \gls{lr} with Tikhonov regularization with $\lambda=1$, training each class against all others.
    \item \emph{\gls{qda}}~\cite{rao1948utilization}: this method attempts to fit a Gaussian density function to each class to obtain a quadratic boundary between classes, which coincides with the maximum likelihood boundary if the data is actually Gaussian.
    \item \emph{\gls{rf}}~\cite{heath1993k}: this model attempts to fit $100$ decision trees on random subsets of the data and input features, using the Gini coefficient as a splitting criterion and applying majority voting over the output.
    \item \emph{Extra Trees}~\cite{geurts2006extremely}: this method is similar to \gls{rf}, but uses $800$ trees, which randomly sample possible splitting points and select the best one, instead of using the Gini coefficient to directly compute the optimal splitting. This increases the randomness of the method.
    \item \emph{\gls{lgbm}}: this model is also based on decision trees, using gradient boosting~\cite{bentejac2021comparative} over $100$ of them to improve ensemble performance. Additionally, \gls{lgbm} includes some additional optimizations over the gradient for faster computation.
    \item \emph{Ensemble}: we consider a soft voting method over the $5$ classifiers above, averaging the predicted probabilities of each class. The output is the most likely user, weighted by the confidence of the models.
\end{itemize}

\section{Experimental results}\label{sec:results}

We consider the performance of the learning methods on all the games in the dataset, using the standard accuracy metric:
\begin{equation}
    A=\frac{\sum_{i=1}^N\delta(y_i,\hat{y}(x_i))}{N},
\end{equation}
where $y_i$ is the actual user index, $\hat{y}(x_i)$ is the index predicted by the classifier, and $\delta(m,n)$ is the Kronecker delta function, equal to $1$ if the inputs are identical and $0$ otherwise. The accuracy is computed over the test set, i.e., the section of each trace from the $8$ minute mark to the $10$ minute mark.
We also consider the macro F1 score for each classifier, which is computed as
\begin{equation}
    F_1=\sum_{y\in\mathcal{Y}}\frac{2\sum_{i=1}^N\delta(\hat{y}(x_i),y)\delta(y_i,y)}{\sum_{i=1}^N\left(\delta(\hat{y}(x_i),y)+1\right)\delta(y_i,y)+\delta(\hat{y}(x_i),y)(1-\delta(y_i,y))}.
\end{equation}

The overall results are reported in Table~\ref{tab:accuracy}. 
We did not report scores for inter-game identification performance, as no learning model obtained an accuracy higher than $0.3$: movement and traffic patterns are game-specific and do not translate directly from one game to the other.
We also tested a game recognition classifier, using the same models described for identification, but we did not report the corresponding performance in Table~\ref{tab:accuracy} since we obtained $100\%$ accuracy on the test set.
  
Considering movement analysis, the accuracy is significantly lower for \emph{Medal of Honor} and \emph{Cooking Simulator}, around $80\%$ for the best models. On the other hand, \emph{Beat Saber} and \emph{Forklift Simulator} have an identification performance higher than $95\%$. We can partly explain this gap with the features of the games. 
The users in \emph{Medal of Honor} and \emph{Cooking Simulator} significantly employed teleporting based on controller button inputs for in-game movements, while this did not occur for \emph{Beat Saber} and \emph{Forklift Simulator}. This difference arises because the former two games involve navigating spaces much larger than the laboratory used for the experiments, as reported in~\cite{baldoni2024questset}. As a result, real-world movements are more accurately mapped in the in-game movements for the latter two.

In \emph{Foklift Simulator}, users played sitting down and performed a driving activity. This implies that users performed repetitive movements whose amplitude depended on their physical characteristics (e.g., arm length) and driving style. In \emph{Beat Saber}, the users were standing still while slashing virtual cubes. Therefore, users' motions were repetitive and highly dependent on users' physical features and agility, making their identification easier. In general, Extra Trees and \gls{rf} are the best models for movement analysis, while the performance of \gls{qda} is particularly low due to the large number of features. Since the algorithm fits Gaussian distributions with arbitrary covariance matrices to the classes, it can easily overfit, obtaining perfect accuracy on the training set but a very low performance over the test set. Finally, combining the models in an ensemble provides a significant enhancement in performance for \emph{Medal of Honor}, while the best individual model performs better for the other games.

\begin{figure*}[ht!]
\centering
\subfloat[Movements (with normalized height)]
	{
        \includegraphics[width=0.36\linewidth]{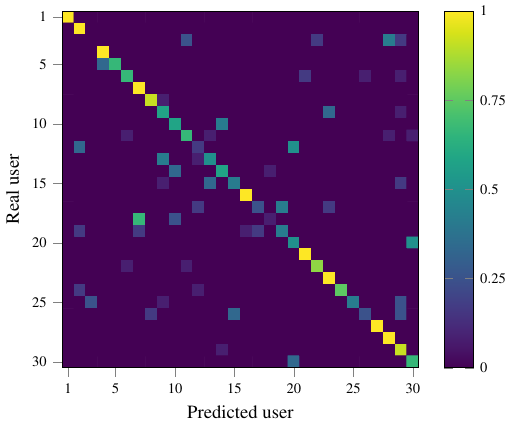}\label{fig:movements_mh}
	}\hspace{1cm}
\subfloat[Movements (with normalized height) and traffic]
	{
        \includegraphics[width=0.36\linewidth]{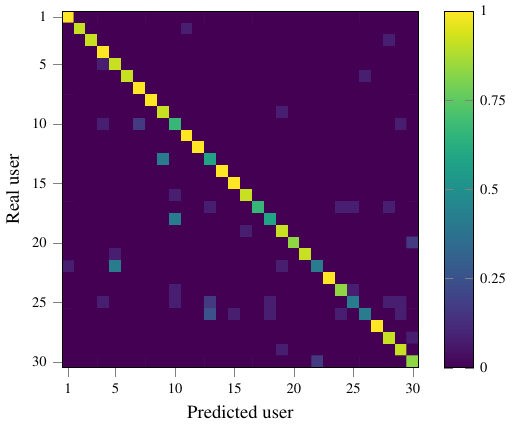}\label{fig:movements_traffic_mh}
	}
    \caption{Confusion matrix for Medal of Honor using only movements and movements and traffic for the \gls{lgbm} classifier.}
    \label{fig:mh}
  \end{figure*}

\begin{figure*}[ht!]
     \centering
     \subfloat[Beat Saber]{\textbf{\includegraphics[width=0.36\linewidth]{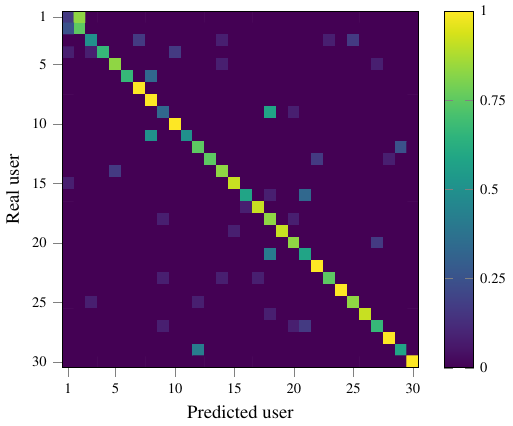}\label{fig:traffic_bs}}}\hspace{1cm}
     \subfloat[Medal of Honor]{\includegraphics[width=0.36\linewidth]{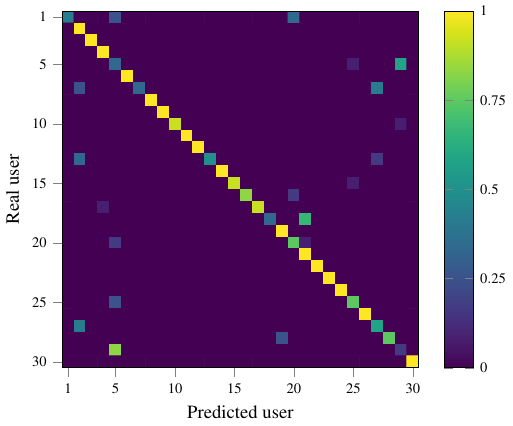}\label{fig:traffic_mh}} \\
     \subfloat[Cooking Simulator]{\includegraphics[width=0.36\linewidth]{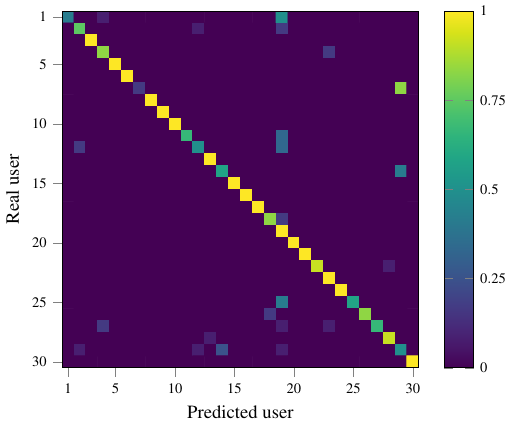}\label{fig:traffic_cs}}\hspace{1cm}
     \subfloat[Forklift Simulator]{\includegraphics[width=0.36\linewidth]{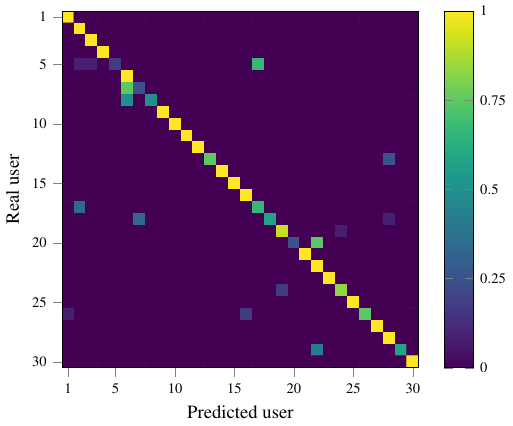}\label{fig:traffic_fs}}
     \caption{Confusion matrices for all games using only traffic with the best performing models.}
     \label{fig:traffic}
\end{figure*}
The game features also explain the results of the performed ablation study. \emph{Forklift Simulator} and \emph{Cooking Simulator} have been categorized in~\cite{baldoni2024questset} as ``slow games,'' whereas \emph{Beat Saber} and \emph{Medal of Honor} have been classified as ``fast games.'' For the first category of games, the identification accuracy is extremely robust: after normalizing the height variables, the decrease ranges between $0\%$ and $3\%$. For fast games, on the contrary, the accuracy drop is between $2\%$ and $11\%$. 
Normalizing with respect to the mean height may have a greater impact on highly dynamic movements, while only slightly affecting the slow ones, which can use other features as biometrics (e.g., the distance between head and hands). We also note that \gls{lgbm} is the most accurate model when normalizing the user height, while the \gls{qda} overfitting issues are not resolved. On the other hand, the ensemble model improves over the best individual algorithm only for \emph{Medal of Honor} and \emph{Cooking Simulator}, indicating that the algorithms make similar errors over the same samples for \emph{Beat Saber} and \emph{Forklift Simulator}.

If we consider traffic-based identification, the accuracy of the best performing classifier is over $80\%$ for all games except \emph{Beat Saber}, which has slightly lower perfomance, but passes this threshold when considering the ensemble model. In the case of \emph{Medal of Honor} and \emph{Cooking Simulator}, traffic-based fingerprinting is more effective than movement analysis. Our hypothesis is that the amount of traffic depends on the game content which, in turn, is influenced by the user behavior. For this reason, traffic data also represents a valuable identification source.  However, the models that perform best are significantly different: \gls{qda} is the best model for all games except \emph{Beat Saber}, as the lower number of features mitigates its overfitting issues, and its complex model captures non-linear features that the other models fail to identify. 

Finally, we can consider a combined identification, including both traffic and movement data: the performance does not significantly improve for \emph{Forklift Simulator}, which already had a high identification accuracy from movement data alone, but it does for the other games. In particular, the identification accuracy for \emph{Medal of Honor} improves by about $12\%$, which increases to $18\%$ when considering normalized height. 
Figure~\ref{fig:mh} shows this improvement on the confusion matrices, considering the case with normalized height. Many of the users that are hard to identify based on movements alone become almost perfectly identifiable if we include traffic data, and identification is uncertain for only a small number of the $30$ total users.

The F1 scores for the best performing models for each game are generally consistent with the accuracy values, with a difference below $2\%$ in almost all games: this indicates that the performance is relatively consistent across different users. 

Our results highlight that movement tracking and traffic data can be effectively employed for user identification, even with simple processing methods. If, on the one hand, this can be used for a seamless and automatic customization of the \gls{vr} experience, on the other, it represents a relevant threat to users' privacy. Using standard network sniffing techniques, an attacker could easily guess the user of the \gls{vr} headset. This is further confirmed by Figure~\ref{fig:traffic}: even when considering \emph{Beat Saber}, i.e., the game with the lowest accuracy when considering traffic data, many users can be identified with almost perfect accuracy. An additional criticality is represented by the fact that the correct classifications are also associated with a high certainty of the classifiers: the attacker is then able to gauge the accuracy of identification attempts as well, obtaining further information.

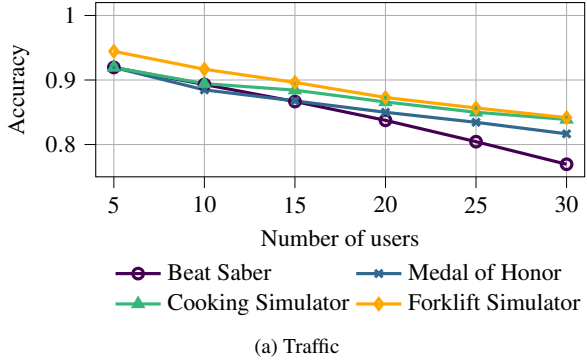
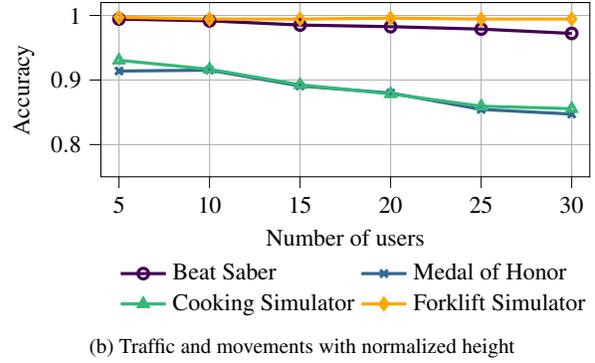
\begin{figure*}[ht]
    \centering
	\subfloat[Traffic]
	{
        \label{fig:trf_user}
\begin{tikzpicture}
\definecolor{darkgray176}{RGB}{176,176,176}
\begin{axis}[
width = \textwidth/2.2,
height = 3.9cm,
legend columns = 2,
legend cell align={left},
legend style={
  fill opacity=0.8,
  draw opacity=1,
  text opacity=1,
  at={(0.02,-0.85)},
  anchor=south west,
  draw=none
},
tick pos=left,
tick align=outside,
x grid style={darkgray176},
xlabel={Number of users},
xmajorgrids,
xmin=4,
xmax=31,
xtick style={color=black},
y grid style={darkgray176},
ylabel={Accuracy},
ymajorgrids,
ymin=0.75,
ymax=1.02,
ytick style={color=black}
]

\addplot [very thick, color1, mark=o]
table[row sep=crcr] {%
5 0.91944444\\
10  0.89305556\\
15  0.86666667\\
20  0.8375\\
25  0.80444444\\
30  0.76944444\\
};
\addlegendentry{Beat Saber};
\addplot [very thick, color2, mark=x]
table[row sep=crcr] {%
5 0.91944444\\
10  0.88472222\\
15  0.86759259\\
20  0.85\\
25  0.83444444\\
30  0.81666667\\
};
\addlegendentry{Medal of Honor};
\addplot [very thick, color3,mark=triangle]
table[row sep=crcr] {%
5   0.91944444\\
10  0.89444444\\
15  0.88425926\\
20  0.86597222\\
25  0.85\\
30  0.83888889\\
};
\addlegendentry{Cooking Simulator};
\addplot [very thick, color4,mark=diamond]
table[row sep=crcr] {%
5   0.94444444\\
10  0.91666667\\
15  0.8962963\\
20  0.87291667\\
25  0.85666667\\
30  0.84166667\\
};
\addlegendentry{Forklift Simulator};

\end{axis}

\end{tikzpicture}
	}\hspace{0.2cm}
    \subfloat[Traffic and movements with normalized height]
	{
	    \label{fig:mov_trf_user}
\begin{tikzpicture}

\definecolor{darkgray176}{RGB}{176,176,176}
\definecolor{darkorange2551490}{RGB}{1,102,94}
\definecolor{limegreen018569}{RGB}{215,48,49}
\definecolor{teal1293165}{RGB}{69,117,180}
\definecolor{byzantine}{rgb}{0.74, 0.2, 0.64}

\begin{axis}[
width = \textwidth/2.2,
height = 3.9cm,
legend columns = 2,
legend cell align={left},
legend style={
  fill opacity=0.8,
  draw opacity=1,
  text opacity=1,
  at={(0.02,-0.85)},
  anchor=south west,
  draw=none
},
tick pos=left,
tick align=outside,
x grid style={darkgray176},
xlabel={Number of users},
xmajorgrids,
xmin=4, xmax=31,
xtick style={color=black},
y grid style={darkgray176},
ylabel={Accuracy},
ymajorgrids,
ymin=0.75, ymax=1.02,
ytick style={color=black}
]

\addplot [very thick, color1, mark=o]
table[row sep=crcr] {%
5   0.99444444\\
10  0.99166667\\
15  0.98518519\\
20  0.98263889\\
25  0.97888889\\
30  0.97222222\\
};
\addlegendentry{Beat Saber}
\addplot [very thick, color2, mark=x]
table[row sep=crcr] {%
5   0.91388889\\
10  0.91527778\\
15  0.89074074\\
20  0.87986111\\
25  0.85444444\\
30  0.84722222\\
};
\addlegendentry{Medal of Honor}
\addplot [very thick, color3,mark=triangle]
table[row sep=crcr] {%
5   0.93055556\\
10  0.91666667\\
15  0.89259259\\
20  0.87847222\\
25  0.85944444\\
30  0.85555556\\
};
\addlegendentry{Cooking Simulator}
\addplot [very thick, color4,mark=diamond]
table[row sep=crcr] {%
5   0.99722222\\
10  0.99444444\\
15  0.99444444\\
20  0.99583333\\
25  0.99444444\\
30  0.99444444\\
};
\addlegendentry{Forklift Simulator}

\end{axis}

\end{tikzpicture}
	}
    \caption{Impact of the number of users on the identification accuracy. The best individual model was selected for each game. The results for fewer than $30$ users are averaged over $6$ possible subgroups, shuffling units of $5$ users.}
    \label{fig:user_number}
\end{figure*}

\begin{figure*}[ht]
    \centering
	\subfloat[Traffic]
	{
        \label{fig:trf_voting}
\begin{tikzpicture}

\definecolor{darkgray176}{RGB}{176,176,176}
\definecolor{darkorange2551490}{RGB}{1,102,94}
\definecolor{limegreen018569}{RGB}{215,48,49}
\definecolor{teal1293165}{RGB}{69,117,180}
\definecolor{byzantine}{rgb}{0.74, 0.2, 0.64}

\begin{axis}[
width = \textwidth/2.2,
height = 3.9cm,
legend columns = 2,
legend cell align={left},
legend style={
  fill opacity=0.8,
  draw opacity=1,
  text opacity=1,
  at={(0.02,-0.9)},
  anchor=south west,
  draw=none
},
tick pos=left,
tick align=outside,
x grid style={darkgray176},
xlabel={Number of voting windows},
xticklabels={},
extra x ticks={1,3,5,7,9,11},
xmajorgrids,
xmin=0.5, xmax=11.5,
xtick style={color=black},
y grid style={darkgray176},
ylabel={Accuracy},
ymajorgrids,
ymin=0.75, ymax=1.02,
ytick style={color=black}
]

\addplot [very thick, color1, mark=o]
table[row sep=crcr] {%
1  0.775758\\
3  0.822222\\
5  0.857143\\
7  0.873333\\
9  0.900000\\
11 0.900000\\
};
\addlegendentry{Beat Saber}
\addplot [very thick, color2, mark=x]
table[row sep=crcr] {%
1   0.830303\\
3  0.848148\\
5  0.842857\\
7  0.840000\\
9  0.855556\\
11  0.866667\\
};
\addlegendentry{Medal of Honor}
\addplot [very thick, color3,mark=triangle]
table[row sep=crcr] {%
1   0.863636\\
3  0.900000\\
5  0.909524\\
7  0.933333\\
9  0.955556\\
11  0.966667\\
};
\addlegendentry{Cooking Simulator}
\addplot [very thick, color4,mark=diamond]
table[row sep=crcr] {%
1   0.851515\\
3  0.851852\\
5  0.852381\\
7  0.860000\\
9  0.855556\\
11  0.900000\\
};
\addlegendentry{Forklift Simulator}

\end{axis}
\end{tikzpicture}
	} \hspace{0.2cm}
    \subfloat[Traffic and movements with normalized height]
	{
	    \label{fig:mov_trf_voting}
\begin{tikzpicture}

\definecolor{darkgray176}{RGB}{176,176,176}
\definecolor{darkorange2551490}{RGB}{1,102,94}
\definecolor{limegreen018569}{RGB}{215,48,49}
\definecolor{teal1293165}{RGB}{69,117,180}
\definecolor{byzantine}{rgb}{0.74, 0.2, 0.64}

\begin{axis}[
width = \textwidth/2.2,
height = 3.9cm,
legend columns = 2,
legend cell align={left},
legend style={
  fill opacity=0.8,
  draw opacity=1,
  text opacity=1,
  at={(0.02,-0.9)},
  anchor=south west,
  draw=none
},
tick pos=left,
tick align=outside,
x grid style={darkgray176},
xlabel={Number of voting windows},
xticklabels={},
extra x ticks={1,3,5,7,9,11},
xmajorgrids,
xmin=0.5, xmax=11.5,
xtick style={color=black},
y grid style={darkgray176},
ylabel={Accuracy},
ymajorgrids,
ymin=0.75, ymax=1.02,
ytick style={color=black}
]

\addplot [very thick, color1, mark=o]
table[row sep=crcr] {%
1   0.975758\\
3  0.977778\\
5  0.980952\\
7  1\\
9  1\\
11  1\\
};
\addlegendentry{Beat Saber}
\addplot [very thick, color2, mark=x]
table[row sep=crcr] {%
1   0.842424\\
3  0.900000\\
5  0.966667\\
7  0.933333\\
9  0.944444\\
11  0.966667\\
};
\addlegendentry{Medal of Honor}
\addplot [very thick, color3,mark=triangle]
table[row sep=crcr] {%
1   0.857576\\
3  0.911111\\
5  0.938095\\
7  0.973333\\
9  1\\
11  1\\
};
\addlegendentry{Cooking Simulator}
\addplot [very thick, color4,mark=diamond]
table[row sep=crcr] {%
1   0.993939\\
3  1\\
5  1\\
7  1\\
9  1\\
11  1\\
};
\addlegendentry{Forklift Simulator}

\end{axis}
\end{tikzpicture}
	}
    \caption{Performance with varying number of voting windows. The best individual model was selected for each game. Performance is averaged over a rolling window.}
    \label{fig:voting_window}
\end{figure*}
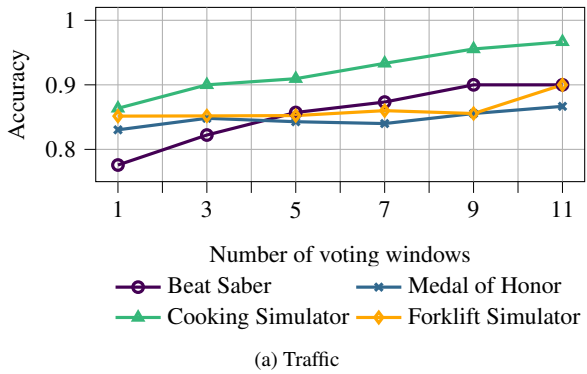
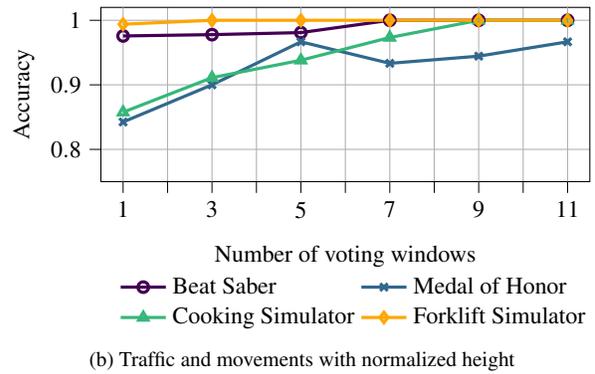

\section{Threats and opportunities}\label{sec:threats}
In the following, we analyze the potential impacts of the results presented in Section~\ref{sec:results} on security aspects, i.e., both on the opportunities for user-friendly identification and on the threats posed by traffic analysis of Cloud \gls{vr}. We also present additional analyses aimed at gauging the effectiveness of the approach in different contexts, such as a different user pool or an attack over a longer time period. In addition, we consider the importance of the features for different games. This helps to better understand the components that have the highest relevance and that future research might focus on to improve identification performance. \edit{We note that our analysis involves the \textit{identification} of users, and not their \textit{authentication}: in the former, which we have analyzed in this study, the problem involves selecting the likeliest member of a set of known users, while in the latter, there is an additional ``intruder'' option that should be selected if the user does not belong to the set of legitimate users. This distinction and the possible extensions of our model to include authentication are discussed in detail in Sec.~\ref{sec:limits}.}

Firstly, we remark that the identification performance achieved by the models considering both movements and traffic, even when normalizing height, provides a solid foundation for simple and user-friendly user identification. The high accuracy in identification for games such as \textit{Beat Saber} or \textit{Forklift Simulator} allows the implementation of passwordless access in low-stakes applications (e.g., access to game libraries and personalized content in home entertainment systems), or provides an additional authentication factor in higher-stakes ones, such as industrial teleoperation systems. Moreover, subtle movement patterns are extremely hard to mimic for other humans, and the related traffic patterns are hard to fake convincingly, thus highlighting the opportunity in using this identification approach.
On the other hand, the accuracy of purely traffic-based re-identification is a relevant threat: the identification of users by a provider or eavesdropper in the network is both a significant privacy issue for individual \gls{vr} users and a security threat for institutional users such as companies or governments, as being able to identify individual operators may lead the attackers to infer confidential information. In the following, we will analyze whether changing the size of the user pool or the observation time can affect these basic considerations, and provide further insight on potential applications.

\begin{table*}[htb]\caption{Most important features across the games \edit{with the corresponding Shapley values. For each game, the first feature is reported in bold and underlined, the second is shown in bold, and the third feature is underlined}.}
\centering
\scriptsize{\begin{tabular}{c|l|c|c|c|c}
\toprule
Feature ID & Feature name      & Beat Saber & Forklift Simulator & Medal of Honor & Cooking Simulator
                                \\ \midrule
1              & Bitrate ($50$-th percentile)   & \underline{$\mathbf{1.81\times 10^{-18}}$} & $3.61\times 10^{-18}$ & $0.03$&  $\mathbf{0.05}$     \\
2              & Number of packets in uplink      & \underline{$1.65\times10^{-18}$}  & $6.42\times10^{-18}$  & $0.05$ & \underline{$\mathbf{0.07}$}   \\
3              & Bitrate ($25$-th percentile)    & $\mathbf{1.81\times 10^{-18}}$ & $3.99\times10^{-18}$&  \underline{$\mathbf{0.08}$} & $0.01$                \\
4              & $z$ (left-right) coord. of head position ($25$-th percentile)     &    $0.56\times10^{-18}$  & \underline{$\mathbf{13.55\times10^{-18}}$}&       $0.06$ & $0.04$            \\
5              & $x$ (close-far) coord. of left controller position (mean) & $0.73\times10^{-18}$ & $\mathbf{8.03\times10^{-18}}$ & $0.02$ & $0.01$ \\
6              & Number of packets in downlink & $0.78\times10^{-18}$ & $2.04\times10^{-18}$ & $\mathbf{0.07}$ & $0.05$  \\
7              & $W$ component of head orientation ($25$-th perc.)    & $1.45\times10^{-18}$ & \underline{$7.00\times10^{-18}$} & $0.01$ & $0.01$\\
8              & Bitrate (max) & $0.85\times10^{-18}$ & $3.44\times10^{-18}$& \underline{$0.07$} & $0.02$ \\
9              & Bitrate (mean) & $1.38\times 10^{-18}$  & $1.06\times10^{-18}$ & $0.02$& \underline{$0.05$}                                            \\ \bottomrule
\end{tabular}\label{tab:feature_names}}
\end{table*}

\subsection{Impact of the time window and number of users}



We analyzed the user identification performance by varying the number of users and enlarging the considered time window both
using only traffic for user re-identification, and 
using traffic and movements with normalized height. 

Concerning the first issue, we evaluated the accuracy of the best-performing classifier for each game by selecting subsets of users. The results were obtained by considering units of $5$ users, which were then combined into $6$ possible groups of $10$, $15$, $20$, and $25$ users, over which the identification performance was averaged using a procedure similar to $K$-fold cross-validation.  It is important to underline that in typical \gls{vr} application scenarios, such as home entertainment and education, the number of users usually ranges around $5$ and $25$, respectively.
Overall, the performance decreases when the number of users increases, as shown in Figure~\ref{fig:user_number}. \edit{When we consider traffic data exclusively,} 
Figure~\subref*{fig:trf_user} shows that the performance drop is approximately linear but relatively slow, with an accuracy loss below $10\%$ for all games except \textit{Medal of Honor} when going from $5$ to $30$ users, highlighting that the user identification through the traffic-based behavioral fingerprint is robust with respect to the number of users in realistic scenarios.

Concerning 
user identification \edit{combined with movements with normalized height}, whose performance is shown in Figure~\subref*{fig:mov_trf_user}, it is possible to notice that 
the game content and the relation between physical and virtual movements also play a relevant role. In more detail, the drop in performance is almost zero for games in which the movements in the virtual world resemble the physical ones (i.e., \emph{Beat Saber} and \emph{Forklift Simulator}). In contrast, when the real actions do not match completely the virtual ones, the performance decrease is larger. The rationale behind this behavior could be that while for \emph{Beat Saber} and \emph{Forklift Simulator} the variability between users is fully expressed by physical movements, part of this information is lost with \emph{Cooking Simulator} and \emph{Medal of Honor}, for which in-game and real movements do not match. 


\begin{figure}[t]
    \centering
    \includegraphics[width=\columnwidth]{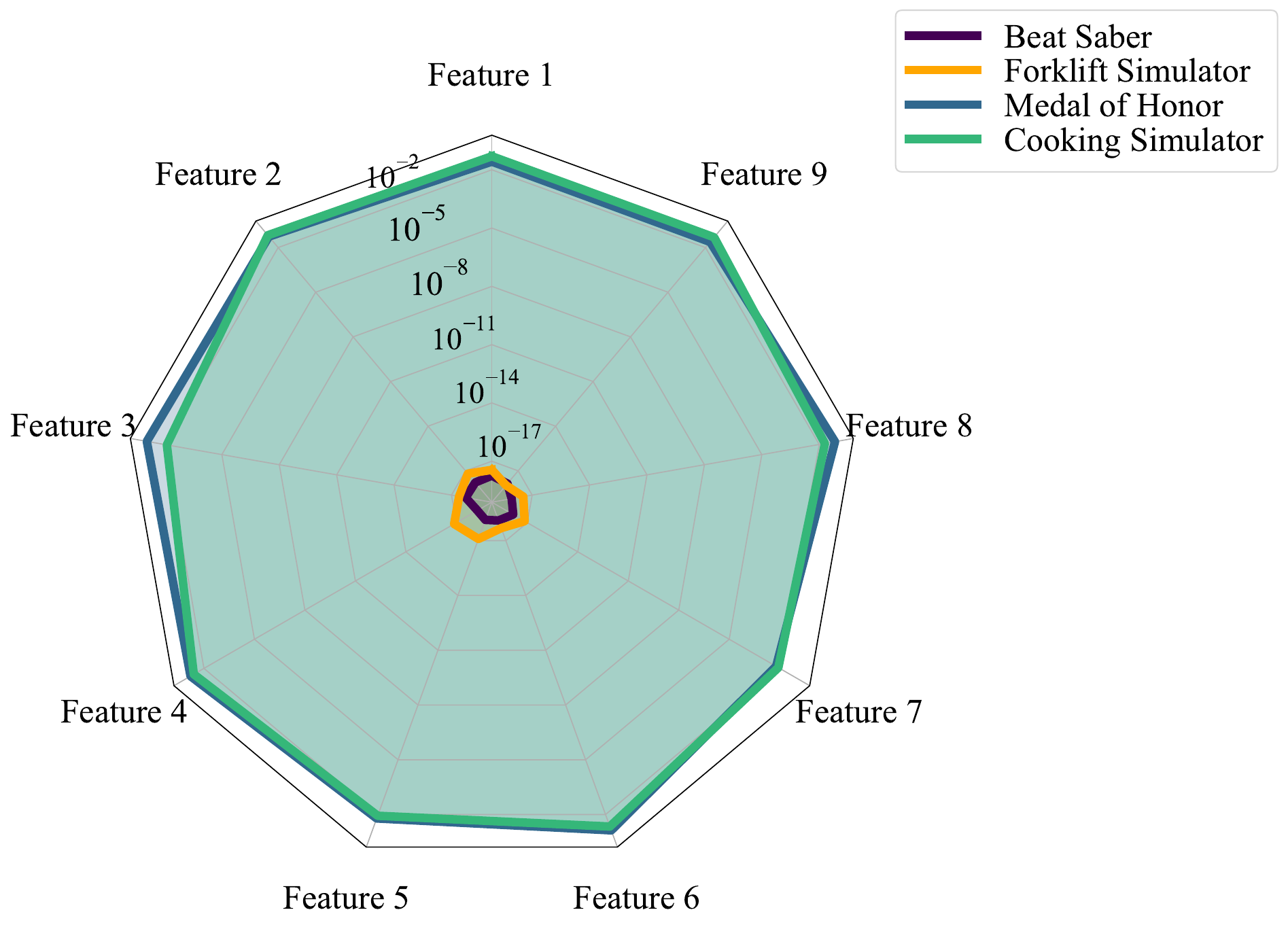}
    \caption{Feature importance based on the Shapley values for the different games, using traffic and movements with normalized height. The best performing individual classifier has been considered for each game.} 
    \label{fig:feat_selection}
\end{figure}

As to the impact of the considered time window, we performed the identification procedure by expanding the identification duration. More specifically, on the threat side, we mimicked an attacker who tries to re-identify the user from the first $10$ seconds of the testing set, and then performs multiple classification attempts on the following slots of $10$ seconds, applying a majority voting strategy. This procedure has been applied for the duration of the entire test set  (i.e., $2$ minutes), considering odd sets of time slots, thus resulting in voting windows with up to $11$ attempts. The results are reported in Figure~\subref*{fig:trf_voting}. For all games, the accuracy tends to increase with a longer voting window, improving accuracy significantly for the traffic-based analysis in \textit{Cooking Simulator} and \textit{Beat Saber}. On the other hand, the improvement is limited for the other two games: this could indicate an inherent difference in the games, but it might be due to a few outliers due to the limited number of available samples in the test set. It is interesting to note that this could be exploited by an attacker, since the accuracy reaches $90\%$ for all games except \textit{Medal of Honor} in the most convenient scenario. Therefore, an optimal voting window could be identified by performing multiple test trials, improving the effectiveness of the attack.
Concerning the case in which both traffic and movement data are used, Figure~\subref*{fig:mov_trf_voting} shows that observing users over a longer time window is beneficial. In more detail, all games achieve an accuracy higher than $95\%$ by increasing the number of samples in the voting window.
It is also interesting to note that, with the exception of \textit{Medal of Honor}, all games reach almost $100\%$ accuracy after $90$~seconds, lending credibility to the option of passwordless identification even for high security applications.


\subsection{Feature importance}\label{sec:feat_importance}

To evaluate the influence of the different features on the identification performance, we computed the average Shapley values~\cite{Fryer_Access_2021} for the 30-user classification problem, focusing on the best-performing classifiers for each game. To analyze the impact of both traffic and movements' characteristics, we performed this analysis for the identification procedure involving both traffic and movement data. 

Figure~\ref{fig:feat_selection} reports the three features which achieved the highest score for the four games. The different features are reported in Table~\ref{tab:feature_names}, together with the corresponding \edit{Shapley values}.

It is interesting to note that the features that play a more relevant role in the classification are extracted both from movements and from traffic. This highlights that both sources are informative for user identification and confirms that it is possible to extract an overall biometric fingerprint considering their joint processing. Concerning traffic data, the most relevant features are associated with the bitrate and the amount of traffic.  
This may confirm our hypothesis that predictive compression and resource allocation techniques are based on the specific actions of the user, so that they provide a useful contribution for user identification. As to movement data, the selected features concern the head position and orientation, as well as the left controller position. 

For different games, the classifier uses different features. For \emph{Beat Saber}, the first three features are extracted from traffic. This indicates that, since this game is highly dynamic, the bitrate and the number of packets reflect well the playing style of each user. In contrast, the slow gameplay of \emph{Forklift Simulator} led the classifier to focus more on movement characteristics.
In \emph{Cooking Simulator} and \emph{Medal of Honor}, the three most important features are extracted from traffic, since physical movements only partially match in-game actions. 
Anyway, all the considered features have similar importance given their similar Shapley values, as shown in Figure~\ref{fig:feat_selection}.
Finally, it is interesting to note that the games can be split in two groups based on the magnitude of feature relevance: i) \emph{Beat Saber} and \emph{Forklift Simulator} and ii) \emph{Medal of Honor} and \emph{Cooking Simulator}.  This can be explained by the fact that in the former case in-game actions resembled the physical ones, while in the latter the two were not matching. 

\subsection{Inter-activity identification}

To analyze the content generalization capabilities of the user identification, we exploited the fact that in Questset each user played two games. More specifically, the games were split into two groups (fast and slow), and each participant tried one game from each category. Therefore,  we trained each classifier on one of the games (fast or slow) and tested on the other (slow or fast). As we mentioned previously, in this case the performance dropped below $30\%$ for all games and feature sets. This outcome is expected: as reported in~\cite{baldoni2024questset}, the games in the dataset were selected specifically for their different contents and movement features. This is further confirmed by the analysis performed in Section~\ref{sec:feat_importance}, where we show that the most important classification features are different for each game. 
This reduces the flexibility of identification across multiple applications, as the considered classifiers need to be trained on the specific game or application to work. On the other hand, it reduces the attack surface with respect to traffic sniffers: in order to identify the user of a \gls{vr} game, the attacker would need to have at least a few minutes of traffic from that same user to train the classifiers. Moreover, before applying a classifier the attacker should identify the current game. Although this can be easily achieved by analyzing the traffic traces, as mentioned in Section~\ref{sec:results}, it represents an additional step that the potential attacker must perform before identification.


\section{\edit{Limitations and future extensions}}\label{sec:limits}

\edit{This work is a first step towards understanding whether the combination between network traffic and movements patterns is suitable for user identification in \gls{vr}. The achieved results highlight the opportunities in terms of customization deriving from automatic user identification while pointing out relevant privacy concerns. However, the current study has some limitations, which must be considered when applying and interpreting its results and also suggest some interesting extensions.} 

\edit{The first natural extension of this work concerns the dataset: each game included $30$ users and a single session, making identification easier. While similar conditions may occur when the identification is performed over a relatively small subset of people, e.g., a class of students or a group of remote control operators for a specific machine, considering datasets including more users, as well as multiple sessions for each user, would improve the robustness of the results. Additionally, the classical classifier models we adopted provide good results but appear to be game-specific, thus being limited in terms of cross-task generalization. The inter-session identification and the generalization across multiple applications will be investigated in future works by relying on more complex deep learning models.}

\edit{Secondly, the analysis focuses on user \emph{identification}, with a limited applicability to authentication systems: considering the possibility of external attackers who might try to impersonate users is another natural extension of the work. The identification approach is also limited to cases in which the attacker already knows users' movement profiles, thus future works will consider an unsupervised approach in which the eavesdropper can distinguish different users without any prior experience.}

\edit{Moreover, future contributions will be focused on the definition of a detailed threat model, which has not been addressed in this work. Considering the current trend of immersive systems, which are evolving towards shared and interactive applications, the practical implications of the identified threats are becoming realistic. As highlighted in~\cite{Nair_USENIX_2023}, in multi-user applications, each \gls{vr} headset sends telemetry information to an external server that forwards it to the other users. Moreover, in the case of cloud gaming, the application content will be streamed from the server to all the users involved. This opens to many possible weak points in the \gls{vr} pipeline, going from the software installed in the sending and/or receiving headset, to channel-level and server-level attacks showing that a detailed threat model is needed.}

\edit{Finally, the possible countermeasures to the highlighted threats have to be identified. Although some countermeasures have been proposed in~\cite{Nair_UIST_2023}, they are not directly applicable to the scenario considered in this work. Concerning movement-based identification, the authors of that study propose to include a multiplicative offset to telemetry data so that real physical features (e.g., height and wingspan) are hidden. However, in our work we did not process the telemetry attributes directly, considering the statistics of the position and orientation of head and hands in time. As a final remark, concerning height, it is useful to underline that in the performed ablation study we showed that the proposed approach is still able to perform user identification after removing the height information. Therefore, the offset-based procedure proposed in~\cite{Nair_UIST_2023} might not be sufficient for avoiding user identification. In addition, as highlighted also by the authors in~\cite{Nair_UIST_2023}, the user may switch off the offset-based defenses in games like Beat Saber, since they might have a relevant impact of the user quality of experience and gaming performance. As to the traffic-based identification, the countermeasures proposed in~\cite{Nair_UIST_2023} are not applicable to the considered scenario, since they target the threats identified in~\cite{Nair_POPETS_2023}. A possible countermeasure would be to introduce padding or a buffering strategy in order to make the network packets uniform in size. However, the first option would result in a waste of network resources, while the second could have a heavy impact on the final user experience. Therefore the threat model should also include an analysis of the effectiveness of the countermeasures, depending on the level of awareness of the attacker, from black-box to white-box approaches. These possible countermeasures, together with their impact on network usage and user experience, will be detailed and tested in future contributions.}

\section{Conclusions}

In this work, we investigated the threats and opportunities deriving from user identification in \gls{vr}. On the application side, the possibility of performing automatic customization depending on the user would be a relevant feature for headsets that are shared by multiple users, such as home-entertainment or education systems. We demonstrated that by jointly processing movement and traffic information it is possible to obtain a reliable user fingerprint for four commercial applications, achieving an identification accuracy above $80\%$ on all games, and of about $98\%$ on two of them. Moreover, we proved that the reliance on internet-based functionalities for networked \gls{vr} applications also creates new threats. To prove this assertion we performed user identification from only traffic data achieving a classification accuracy of about $80\%$ on all games. This highlights that security issues should not be underestimated in \gls{vr} applications. 


\acknowledgments{
This work was partially supported by the European Union under the Italian National Recovery and Resilience Plan (NRRP) Mission 4, Component 2, Investment 1.3, CUP C93C22005250001, partnership on “Telecommunications of the Future” (PE00000001 - program “RESTART”)}

\bibliographystyle{abbrv-doi}

\bibliography{paper.bib}
\end{document}